\documentstyle[psfig]{article}

\begin{document}
\title{Quantum Computing Hamiltonian cycles.}
\author{T.Rudolph\\Department of Physics \& Astronomy\\York University\\4700 
Keele St.\\Toronto \\Ontario M3J 1P3\thanks{email: rudolph@helios.sci.yorku.ca}}

\date{29th February 1995.}
\maketitle
\begin{center}
\begin{abstract}
An algorithm for quantum computing Hamiltonian cycles of simple, cubic,
 bipartite graphs is discussed. It is shown that it is possible to
 evolve a quantum computer into an entanglement of states which map onto the
 set of all possible paths initiating from a chosen vertex, and furthermore to
 subsequently project out all states not corresponding to Hamiltonian cycles. 
\end{abstract}
\end{center}

\def\>{\rangle}
\def\<{langle}
A Hamiltonian cycle is a path on a graph which visits each vertx $1 .. n$
 exactly once, returning to the original vertex in the final step. Here we will
 discuss only simple (without loops or multiple edges), cubic (each vertex has
 only 3 edges), bipartite (each black vertex is connected only to white
 vertices) planar graphs. Such a graph is shown in Figure 1(a), with a 
Hamiltonian cycle indicated by the darker edges.

\begin{figure*}
\psfig{figure=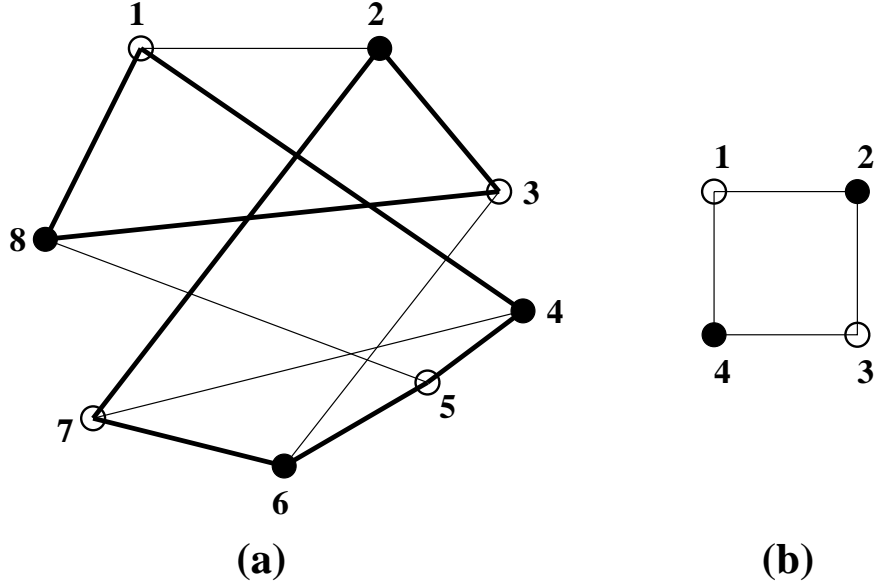}
\caption{(a) A cubic, bipartite graph with Hamiltonian cycle in bold. (b)A simple 2-regular bipartite graph.}
\end{figure*}

Classically no efficient algorithm exists to resolve the question of whether 
any given such graph has a Hamiltonian cycle, although over the years many 
results in Graph Theory have isolated certain special cases. The fundamentals 
of Graph Theory and many associated classical algorithms are well explained in
 \cite{gibb}, and a comprehensive exposition of Hamiltonian cycles and the
 related Travelling Salesman Problem can be found in \cite{law}. A good summary 
of quantum computing can be found in \cite{ekert}.

We shall see that the added power afforded us by a quantum computer's ability to 
carry through parallel computations in a single step enables us to compute all 
possible paths on a given graph. To achieve this we require $n$ registers each 
composed of $n$ qubits, each qubit corresponding to one vertex. Three qubits of 
the first register, denoted $\alpha$, are  involved in every step of the 
algorithm. The other $n-1$ registers contain qubits which in the main are in the
state $|0\rangle$, except for the qubit corresponding to the walkers current 
position.

To clarify the concepts consider the trivial but illustrative case of the square 
in Figure 1(b). We imagine a walker starting at vertex 1 and so prepare our 
quantum computer with register $\alpha$ in the state $|1,0,0,0\rangle_\alpha$,
register 1 in the state $|1,0,0,0\rangle_1\equiv|1\rangle_1$ and the remaining 
registers empty. We envisage a controlled unitary operation $R^j_i$ for the 
$j$th step from vertex $i$. $R^{j}_i$ is conditioned on the $i$th qubit in 
register $j$ being in the state 1. It acts onthe 2 qubits in registers $\alpha$ 
and $j+1$ which correspond to vertices adjacent to $i$.
In a series of three steps we wish the initial state $|1,0,0,0\>_\alpha|1\>_1$
 to evolve as follows (empty registers not shown and register labelling 
dropped),
\begin{eqnarray*} 
R^1_1:&\{ |1,1,0,0\rangle|1\>|2\>; \;|1,0,1,0\rangle|1\>|3\> \} \\
R^2_2 R^2_4:& \{|0,1,0,0\rangle|1\>|2\>|1\>;\; |1,1,1,0\rangle|1\>|2\>|3\>;\; 
|0,0,0,1\rangle|1\>|4\>|1\>; \\ &
|1,0,1,1\rangle|1\>|4\>|3\>\} \\
R^3_1 R^3_3:& \{ |0,0,0,0\rangle|1\>|2\>|1\>|2\>;\; 
|0,1,0,1\rangle|1\>|2\>|1\>|4\>;\;
|1,0,1,0\rangle|1\>|2\>|3\>|2\>;\;\\ & |0,0,0,0\rangle|1\>|4\>|1\>|4\>;\;
|0,1,0,1\rangle|1\>|4\>|1\>|2\>;\; |1,0,1,0\rangle|1\>|4\>|3\>|4\>;\;\\ & 
|1,1,1,1\rangle|1\>|2\>|3\>|4\>;\; |1,1,1,1\rangle|1\>|4\>|3\>|2\> \}.
\end{eqnarray*}   

The \{\}'s denote a superposition of those states enclosed (ignoring 
normalisation and phases for the moment). We see that only those paths 
which are possibly Hamiltonian cycles in the next step contain all 1's 
in register $\alpha$. This is because stepping back to an already passed
 vertex changes a pre-existing 1 at that site to a 0. 
Thus projecting out this state of register $\alpha$ will leave the quantum 
computer in an entanglement of states which correspond to the graph's 
Hamiltonian cycles, if they exist.

\begin{figure*}
\psfig{figure=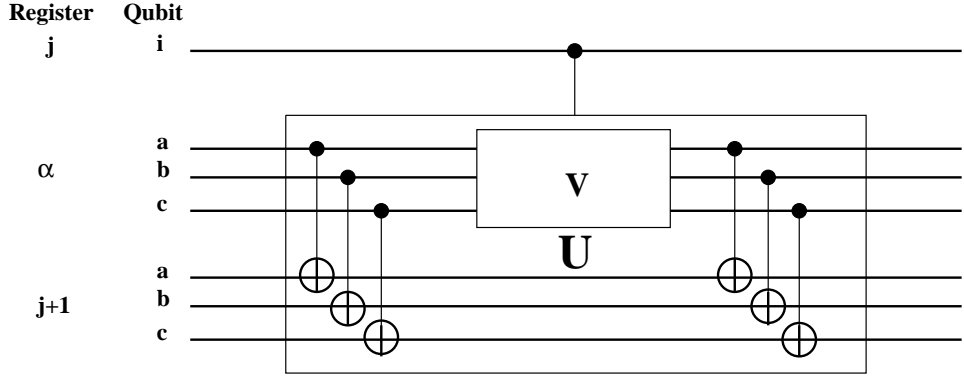}
\caption{A schematic showing a suitable form of operator $U^j_i$, where 
$a,b,c$ label the three vertices adjacent to $i$ in the graph being 
considered.}
\end{figure*}

To progress to the slightly harder problem of cubic graphs we need to be more 
specific about the form of the unitary transform $U^j_i$ required.
It will act on 6 qubits, and also be conditioned on qubit $i$ in register $j$ 
being 1.
 It will apply elementary NOT type operations to the 3 qubits $a,b,c$ in 
register $\alpha$ which label the vertices adjacent to $i$, and write 1's into 
the same sites in the (previously empty) register $j+1$ similarly to 
that discussed above. 
A schematic of such a transform is shown in Figure 2, using 
a version of Feynman's \cite{fey} notation developed in \cite{bar}. The $\oplus$ 
is the elementary (1 bit) NOT operation given by the matrix ${1\,0\choose0\,1}$. 
Time progresses from left to right. The 3 qubit transformation $V$ is given in 
the lexicographically ordered basis $|0,0,0\>;|0,0,1\>,...|1,1,1\>$ by the 
matrix
\begin{equation}
\frac{1}{\sqrt{3}} \left( \begin{array}{cccccccc} 
0&1&1&0&1&0&0&0 \\ 1&0&0&1&0&1&0&0 \\ 1&0&0&1&0&0&1&0 \\ 
0&1&-1&0&0&0&0&1 \\ 1&0&0&0&0&1&-1&0 \\ 0&1&0&0&1&0&0&-1 \\ 
0&0&1&0&1&0&0&1 \\ 0&0&0&1&0&1&1&0 
\end{array} \right).
\end{equation}

  As an example of how the computation would proceed, consider again Figure 1(a). 
A first application of $U^1_1$ would evolve the initial state $|1,0..0\>$ into 
the superposition 
$\frac{1}{\sqrt{3}}(|1,1,0..0\>|1\>|2\>+|1,0,0,1,..0\>|1\>|4\>+|1,0..1\>|1\>|8\>)$. 
Subsequently we would apply $R^2_2R^2_4R^2_8 \equiv R^2_{2,4,8}$ followed by 
$R^3_{1,3,5,7}$, then $R^4_{2,4,6,8}$ and so on until we finally apply 
$R^7{1,3,5,7}$. Projecting out those states which contain all 1's in register 
$\alpha$ will leave the computer in a superposition of states which map 
onto Hamiltonian cycles. A final measurement will reveal one of those cycles, 
should they exist for the particular graph under consideration. 

It is easy to see that in general we require $O(n^2)$ applications of $U^j_i$. 
This is obviously impractical with present technology. However algorithms such as 
the one above help us understand what quantum computers are capable of in 
principle and are therefore important in understanding the relation between 
quantum and classical complexity classes.

\end{document}